\documentstyle{mn}
\title[Emission from Fast Radiative Shocks]
{
Fast Radiative Shocks in Dense Media. III. Properties of the Emission
}
\author
[Tomasz Plewa]
{
Tomasz Plewa\thanks{E-mail address: plewa@sirius.astrouw.edu.pl} \\
Warsaw University Observatory, Al.\ Ujazdowskie 4, 00-478, Warsaw, Poland
}
\date
{
Accepted 1994 .
Received 1994 February 28;
in original form 1994 February 23
}
\volume{000}
\pagerange{000--000}
\begin{document}
\maketitle
\begin{abstract}
Evolution of fast, radiative shocks in high density medium is presented.
Ionizing spectra and approximate broad band light curves of the shocked
gas are calculated. Emergent shock spectra, as seen by a distant
observer, are obtained from photoionization models. The emergent spectra
have a power-law shape $F_{\nu}\propto{\nu}^{-\alpha}$ with mean spectral
index $\alpha\sim0.6-1.0$ in the energy range $0.01-10$ keV, and have a
high-energy cutoff corresponding to the original shock velocity. It is
shown that the models exhibit promising features that may account for
some photometric and spectral properties of Active Galactic Nuclei.
\end{abstract}
\begin{keywords}
hydrodynamics -- shock waves -- instabilities -- ISM: supernova remnants
-- galaxies: active -- X-rays: galaxies
\end{keywords}
\section{Introduction}
In the series of papers Terlevich and his group (see, e.g., Terlevich et
al.\ 1992) developed the starburst model for Active Galactic Nuclei
(AGNs). In this model AGN is powered by compact, dense supernova
remnants (cSNRs), and the bulk of radiation is emitted by the supernova
shock wave evolving in a dense medium ($n_0\sim10^7$ cm$^{-3}$). Terlevich et
al.\ \shortcite{ACEROSI} calculated 1D and 2D hydrodynamical models of
cSNR evolution and demonstrated that it was possible to recover observed
characteristics of Broad Line Region of AGNs. Using very simple
assumptions Terlevich et al.\ (1994a, TTRFM) successfully explained
observed difference between moments of maximum continuum and line emission
taking into account dependence on ionization parameter. While the rapid
X-ray variability of AGNs still remains to be explained in this model,
preliminary investigations \cite{ACEROSIII} showed that interactions
between dense, fast moving clumps of gas could be partially responsible
for observed variations of high-energy emission. Detailed criticism of
the starburst model was presented by Heckman \shortcite{Heck91} and
Filippenko \shortcite{Fil92}.

Apart from their hypothetical presence in AGNs, compact supernova
remnants were detected in nearby galaxies. On the basis of spectroscopic
studies such objects like SN 1980K \cite{SN80K}, SN 1987F \cite{SN87F},
or SN 1988Z (Stathakis \& Sadler 1991; Turatto et al.\ 1993) are
interpreted as supernova remnants evolving in a medium of density as
high as $10^7$ cm$^{-3}$. The lack of appropriate theoretical models was
noted by Leibundgut et al.\ \shortcite{Leib91} in their study of SN
1986J and SN 1980K. Both the possible role of cSNRs in AGNs and the
remark by Leibundgut et al.\ gave motivation to our work on
hydrodynamical evolution of fast shocks evolving in a dense environment.

The subject is not new. Over a decade ago Chevalier \& Imamura
\shortcite{CI82} showed that radiative, steady shocks are subject to an
oscillatory instability. This result was confirmed on the basis of
nonlinear hydrodynamical analysis for nonstationary shocks (Gaetz, Edgar
\& Chevalier 1988; Stone \& Norman 1993) as well as for steady radiative
shocks (Innes, Giddings \& Falle 1987). Both groups found unstable
behavior of shocks faster than $\sim$130 km s$^{-1}$. The spectra of
faster shocks ($v_{\rm s}\le1100$ km s$^{-1}$) were discussed in some
detail by Binette, Dopita \& Tuohy \shortcite{BDT85}. They found that
the emergent shock spectrum can be approximated by a power-law with a
spectral index $\alpha\sim 0.5$ for energies lower than $\sim 1$ keV.
However, all theoretical investigations mentioned above apply to
relatively slow shocks in low density medium ($n<100$ cm$^{-3}$).
Therefore, they are not applicable to conditions expected for cSNRs or
for central regions of AGNs.

In the first paper of the series (Plewa \& R\'o\.zyczka 1992, Paper I)
we investigated the evolution of fast ($v_{\rm s0}\ge1000$ km s$^{-1}$)
radiative shocks in a uniform medium of density $n=10^{7}$ cm$^{-3}$.
The shock and its post-shock region were found to be highly
nonstationary, leading to the conclusion that some stochastic component
might be present in the evolutionary pattern, but we failed to obtain
any definite results due to the low resolution of the models.
Subsequently, we used a high-resolution adaptive grid technique (Plewa
1993, Paper II) and we found quasiperiodic patterns in the evolution of
the shocks. The flow behind the main shock was found to be highly
discontinuous due to the presence of secondary shocks leading to a
stochastic behavior of the shock luminosity.

In Paper I and II our model shocks were formed by the reflection of a
steady stream of gas from a rigid wall, and they were not directly
applicable to any astrophysical situation. In the present paper we
improve the models by allowing for a secular decrease in stream velocity
to mimic the evolution of the forward SN shock. Also, approximate
spectra of shocks are obtained under the assumption of collisional
equilibrium. Description of the method is given in Section
\ref{s:method}. In Section \ref{s:dyn_evol} and \ref{s:ph_and_spec} we
present details of the shock evolution, the broad band photometry and
the spectrum of the radiation emitted by the shock, as well as results
obtained from photoionization models. Possible relation of the models to
AGNs is discussed in Section \ref{s:discussion}.
\section{The method}
						\label{s:method}
The hydrodynamical evolution of the shock is modeled using {\it
Piecewise Parabolic Method} (PPM) of Colella \& Woodward
\shortcite{CW84} in 1D plane-parallel geometry. Our preliminary
investigations presented in Paper I showed that even the high resolution
of the PPM scheme is not sufficient to model accurately the most
luminous and dense regions of the shocked gas if a standard uniform grid
is used. The evolution of such regions is characterized by cooling time
scale much shorter than the dynamical scale of the flow. For example, to
avoid complications arising from the lack of resolution in strongly
radiating regions Blondin, Fryxell \& K\"{o}nigl \shortcite{spectr} used
a {\em special treatment} lowering artificially local emission rates.
Our numerical experiments showed that such procedure strongly affects
the results, and we concluded that to model the evolution correctly a
much higher resolution should be used.

We found that to resolve thin, strongly radiating regions of shocks
faster than $\sim1000$ km s$^{-1}$ the grid should contain about
$10^{4}$ equally spaced zones. The cheapest way to obtain such a high
resolution is the adaptive grid method. In our approach presented in
Paper II the grid motion is governed by an adaptive grid algorithm
developed by Dorfi \& Drury \shortcite{SAGE}. The grid points are
distributed accordingly to a user-defined {\em resolution function}. For
our problem the resolution function was defined as the sum of gradients
of density, internal energy, and emission rate (see Paper II for
details). We also added the term proportional to the gas density to
obtain higher resolution in high density regions. The grid equation was
solved implicitly at the beginning of every hydrodynamical time step.
The grid was divided into 500 zones and we used Courant number equal to
0.8. The number of time steps needed to complete the evolution using
maximum point concentration equal to $2.2\times10^{4}$ was about
$2\times10^{5}$.

At the left boundary of the grid we impose a reflecting boundary
condition, while from the right side the grid is fed with gas of
constant density and temperature. To mimic the decrease of shock
velocity in a way appropriate for SN shock we vary the velocity of the
inflowing gas with time as
\[
v_{\rm in}=v_{0}\times (t/t_{\rm sg}+1)^{-\frac{5}{7}},
\]
where $t_{\rm sg}$ has the interpretation of the time that has to
elapse since the SN explosion before cooling effects become noticeable.
According to Shull (1980)
\[
t_{\rm sg} = 0.63 E_{\rm SN} \left(\frac{n_0}{10^7}\right)^{-3/4}~{\rm yr},
\]
if the explosion energy is $E_{\rm SN}\times10^{51}$ erg and the density
of the ambient medium is $n_0$. This way we account for the fact that in
$\sim t_{\rm sg}$ years after the SN explosion the mean velocity of the
forward SN shock begins to decline like $(t/t_{\rm sg})^{-5/7}$ (Shull
1980; see also Fig.\ 3 in Falle 1981).

This set of initial and boundary conditions leads to the formation of a
strong shock at the left boundary of the grid (Fig.\ \ref{f:sch}).
%
%
\begin{figure}
  \centering
  \vspace{15pc}
  \caption{Schematic representation of initial conditions for shock evolution.}
  \label{f:sch}
\end{figure}
The initial velocity of the shock for constant gamma-law gas with
$\gamma=5/3$ is given by $v_{\rm s0}=\frac{4}{3}v_{0}$. Let us note
that, when compared to real SNR, the left boundary of the grid can be
regarded as the interface between the ejecta and the gas shocked by the
forward supernova shock. Thus, the advantage of our approach is the
possibility to concentrate all grid points in the most interesting
narrow region of the remnant where strong cooling occurs. The
disadvantage is the simplified boundary condition imposed at the left
edge of the grid, which in general may influence the structure and
dynamics of the dense shell. This, however, does not happen in the case
of shock with $v_{\rm s0}\ge3000$ km s$^{-1}$ which form their shells far
away from the left edge of the grid (Fig.\ \ref{f:den}). The reflecting
boundary condition affects only the models of shocks with $v_{\rm
s0}<3000$ km s$^{-1}$, which form two shells, one of them being attached
to the left edge of the grid (Fig.\ \ref{f:den}a). However, as the
latter shell is much less massive than the shell formed close to the
main shock, the influence of the boundary condition on both the dynamics
and luminosity of the post-shock region as a whole is marginal.

Following Terlevich et al.\ (1992) and TTRFM, we assume for the ambient
medium a density $n_{0}=10^7$ cm$^{-3}$ and a temperature $T_{0}=10^4$
K. These values are admittedly arbitrary. However, the choice of
temperature is practically unrestricted because of high shock velocities
(the shock would remain very strong even if $T_0$ was an order of
magnitude higher). As for the density, it was shown in Papers I and II
that its reduction or increase results respectively only in increase or
reduction of the evolutionary time scale and the size of the post-shock
region, while all characteristic evolutionary features of the model
remain unchanged. The adopted value of $n_{0}=10^7$ is particularly
interesting because of indications that it may be characteristic
of the medium surrounding cSNRs.

Radiative losses are allowed for in the post-shock region, while the
shock moves to the right through the yet unshocked gas. The losses are
calculated implicitly assuming collisional equilibrium conditions for
optically thin plasma with solar abundances. We use cooling function
obtained with the help of CLOUDY 84.06 code \cite{cloudy}. We assume
that the heating/cooling balance is achieved at a temperature equal to
the temperature of the unshocked ambient medium $T_{\rm
min}=T_{0}=10^{4}$ K. Therefore, in the regions where temperature has
dropped below $T_{\rm min}$ (e.g.\ due to gas expansion) the temperature
was increased to reach the lower limit given above.

Also under the assumption of collisional equilibrium, spectra of the
radiation emitted by the hot shocked gas are calculated at selected
evolutionary moments. These spectra will be called {\em ionizing
spectra} in order to distinguish them from the {\em emergent shock
spectra} obtained from photoionization calculations presented in Section
\ref{s:contin} below (the latter are the spectra emerging from our
models which would be registered by a distant observer if the space
between the observer and the shock would be empty.) To save the CPU
time, in place of the extremely time consuming CLOUDY code we use the
following approximations for calculation of the ionizing spectrum.
Continuum emission consists of the f-f emission with approximate Gaunt
factor calculated after Hummer \shortcite{Hum88}, while the b-f and
2-photon emissions are calculated using formulae given by Mewe, Lemen \&
van~den Oord \shortcite{MeweVI}. Contribution from over 2600 lines is
calculated using line emissivities taken after Stern, Wang \& Bowyer
\shortcite{Stern78} and Mewe \shortcite{Mewe92} (see also Mewe,
Gronenschild \& van~den Oord 1985).

In order to estimate the relative broad band distribution of energy in
the ionizing spectrum we define a set of three filters corresponding to
wavelength intervals $\lambda\le100$ \AA\ (high energy, HE),
$100<\lambda\le905$ \AA\ (medium energy, ME), and above $905$ \AA\ (low
energy, LE). The filter transmission function is defined as a fraction
of the total energy emitted at a given temperature per given wavelength
interval (LE, ME, HE; Fig.\ \ref{f:fil}).
%
%
  \begin{figure}
  \centering
  \vspace{15pc}
  \caption{Definition of the broad band filters. HE, ME, and LE filter
           transmissions are drawn by thick, thin and dotted lines,
           respectively.}
  \label{f:fil}
\end{figure}
The transmission functions are calculated by detailed integration of the
spectrum emitted by the optically thin plasma as given by the CLOUDY
code.

Crucial to our approach is the assumption of collisional equilibrium
which makes the problem numerically tractable. Admittedly this
assumption does not work well in the lower temperature range. However,
it is justified for the bulk of the shocked gas with temperatures higher
than $5\times 10^{5}$ K (only below that value the cooling function
begins to be history dependent; see, e.g., Shapiro \& Moore 1976; Gaetz
et al.\ 1988). Moreover, as it was demonstrated by Gaetz et al.\
\shortcite{GEC88} details of cooling are not very important for global
shock dynamics (obviously, they play a more significant role in the
evolution of cold regions with $T<10^{5}$ K). The cool regions are the
source of ultraviolet and optical radiation only and we do not expect
significant changes of the high energy emission when nonequilibrium
effects would be considered. These issues are further discussed in
Sect.\ \ref{s:contin}. We do not calculate preionization structure of
the gas upstream from the shock. Also, we give here only results for
solar composition, although higher metallicity would be more appropriate
for both cSNR and AGN applications. Taking all the simplifications into
account, our shock spectra are to be treated as first approximation
only, especially in LE and ME ranges. On the other hand, we are fairly
confident that the dynamics of the shocks has been modeled properly, and
that any improvements in the treatment of cooling processes would not
affect dynamical evolution of the shocks in a significant way.
\section{Dynamical evolution}
						\label{s:dyn_evol}
We calculated models for initial shock velocities equal to 1000, 2000,
4000, and 6000 km s$^{-1}$.  It should be stressed here that only the
fastest shock model may be directly related to cSNRs evolving in a
uniform medium of $n_0 =10^7$ cm$^{-3}$, as, according to Shull (1980),
in such a case the velocity of the SN shock at $t=t_{\rm sg}$ is equal
to $\sim 6000$ km s$^{-1}$. The remaining models have no direct
application, and they only serve to illustrate the strong dependency of
the evolution on $v_{\rm s0}$ (one may expect, however, slower SN shocks
to appear in nonuniform media, e.g., in dense shells of wind bubbles
blown by progenitors.)

Here we present details of the evolution for shocks of velocities equal
to 2000 (the {\em slow} shock) and 6000 km s$^{-1}$ (the {\em fast}
shock), as the evolution in the two remaining cases shows no
qualitatively new features. The evolution was followed up to
evolutionary time $t=3$ yr and $t=12$ yr for slow and fast shock,
respectively. Summary of the results for all the cases considered is
presented in Table \ref{t:basres}. Note that the total shock luminosity
was calculated assuming that the radius of the emitting sphere is equal
to $R_{\rm sh}=3\times10^{16}$ cm (see Fig.\ 8a of Terlevich et
al.\ 1992). Evolution of the model shocks in form of a sequence of
vertically shifted density plots in logarithmic scale is shown in Fig.\
\ref{f:den}
%
%
\begin{figure*}
  \centering
  \vspace*{20pc}
  \caption{Density evolution of model shocks.
           (a) initial shock velocity: $v_{\rm s0}=2000$ km s$^{-1}$;
               distance scale: $1.1\times10^{15}$ cm;
               time scale: 3 years;
               density scale: $1.78\times10^{-18}-2.90\times10^{-13}$ g
cm$^{-3}$;
           (b) initial shock velocity: $v_{\rm s0}=6000$ km s$^{-1}$;
               distance scale: $1.2\times10^{16}$ cm;
               time scale: 12 years;
               density scale: $1.79\times10^{-18}-1.26\times10^{-13}$ g
cm$^{-3}$.
           The vertical scale may be read from the jump in the lowermost
           which corresponds to an increase of density by a factor of 4.
           See text for details.}
  \label{f:den}
\end{figure*}
(all plots were taken at equidistant intervals of time). The left
boundary of the plot corresponds to the contact discontinuity between
the ejecta and the ambient medium, and the ambient gas enters the grid
from the right side.
\subsection{Evolutionary phases}
						\label{s:evol_phases}
In all cases listed in Table \ref{t:basres}
%
%
\begin{table}
  \centering
  \caption{Basic parameters of the model shocks.}
  \label{t:basres}
  \begin{tabular}{cccc}
  \hline \hline
     $v_{\rm s0}$ &   $R_{\rm max}$    &      $F_{\rm max}$      &    $L_{\rm
max}$    \\
      km s$^{-1}$ &        cm          &  erg cm$^{-2}$s$^{-1}$ &    erg
s$^{-1}$    \\
  \hline
        1000       & $1.8\times10^{14}$ &    $7.5\times10^{6}$    &
$9.4\times10^{40}$ \\
        2000       & $1.0\times10^{15}$ &    $3.7\times10^{7}$    &
$4.7\times10^{41}$ \\
        4000       & $4.7\times10^{15}$ &    $8.7\times10^{7}$    &
$1.1\times10^{42}$ \\
        6000       & $1.2\times10^{16}$ &    $1.5\times10^{8}$    &
$1.9\times10^{42}$ \\
  \hline
  \end{tabular}

$v_{\rm s0}$ -- the initial shock velocity; $R_{\rm max}$ -- the maximum
extension of the shock; $F_{\rm max}$ -- the maximum energy flux;
$L_{\rm max}$ -- total shock luminosity assuming the radius of the
emitting sphere equal to $R_{\rm sh}=3\times10^{16}$ cm.
\end{table}
the evolution of the shock could be divided into three distinctive phases:
\begin{enumerate}[{3}]
\renewcommand{\theenumi}{(\arabic{enumi})}
\item {\em Nearly adiabatic shock expansion.}
The evolution starts with a nearly adiabatic shock expansion as cooling
is ineffective at temperatures over $10^{7}$ K for typical cooling
function (see Raymond, Cox \& Smith 1976). As both temperature of
radiating gas and velocity of inflowing gas decrease with time, the
shock slows down while cooling efficiency begins to grow with
temperatures falling below $10^7$ K.
\item {\em Thin shell formation.}
At temperatures lower than $\sim10^{6}$ K the cooling time scale
suddenly drops down. This phenomenon is known as {\em catastrophic
cooling} (Falle 1975, 1981) and it results in formation of thin
transition regions (hereafter referred to as {\em cooling waves}) in
which thermal energy of the gas is rapidly radiated away at nearly
constant density. The cooling waves are the source of almost all the
emission generated by the model in ME and LE energy bands. As our
resolution function contains the term proportional to the luminosity
gradient, cooling waves are well resolved.

Soon after the formation of cooling waves the cool gas begins to
condensate into a very thin and dense shell, visible as the highest peak
in both panels of Fig.\ \ref{f:den} (the less prominent spikes in Fig.\
\ref{f:den} originate from minor flow discontinuities like weak shocks).
At this moment the grid points strongly concentrate in the region of
shell formation (the maximum grid concentration corresponds to
$2.2\times10^{4}$ equally spaced zones). This grid concentration is
close to a practical upper limit, as at significantly higher resolutions
the time step becomes prohibitively short due to the CFL condition. This
phase of the evolution really challenges our adaptive grid method which
has to redistribute the grid points between the main shock profile, the
fast moving cooling waves, and the rapidly growing shell. All these
structures are well resolved, proving the efficiency of the method.
\item {\em Oscillatory instability of the main shock.}
Finally, loss of the pressure in the post-shock region due to cooling
leads to the recession of the main shock, and to its reflection from the
dense shell marked by a kink in shock position in Fig.\ \ref{f:time_dyn},
%
%
\begin{figure*}
  \centering
  \vspace*{15pc}
  \caption{
    Temporal variations of position and velocity of the main shock
    (dashed and solid lines, respectively).
(a) initial shock velocity $v_{\rm s0}=2000$ km s$^{-1}$;
    distance scale: $1.04\times10^{15}$ cm;
    velocity scale: 2000 km s$^{-1}$.
(b) initial shock velocity $v_{\rm s0}=6000$ km s$^{-1}$;
    distance scale: $1.18\times10^{16}$ cm;
    velocity scale: 6000 km s$^{-1}$.
}
  \label{f:time_dyn}
\end{figure*}
in which position and velocity of the shocks are plotted against time.
In Paper I we showed that the velocity of the reflected shock depends on
momentary structure of the shell and parameters of the inflowing gas. At
the moment of the shock reflection the velocity of the inflowing gas is
by a factor of few lower than at the beginning of the evolution. This
implies lower velocities of the reflected shocks ($v_{\rm rs}<1000$ km
s$^{-1}$) and, therefore, shorter evolutionary time scales. The
evolutionary time scale of the main shock is additionally reduced due to
strong cooling in the region just behind the shock, as after the
reflection the temperature of the post-shock gas never exceeds $10^{7}$
K. The mean period of the oscillations is roughly equal to $P_{\rm
osc}\approx14.5$ d and $P_{\rm osc}\approx3.3$ d for fast and slow
shock, respectively. The period of the main shock oscillations
corresponds to the fundamental mode of the oscillatory instability
discussed by Chevalier \& Imamura \shortcite{CI82} and Gaetz et al.\
\shortcite{GEC88}. The oscillatory phase is preceded by a relatively
short period of time during which the density contrast between the
inflowing gas and the shell is too low for the shock reflection to occur
(the reflected shock is very weak and its cooling time is extremely
short). This phase is marked by a chaotic behavior of luminosity and
velocity of the main shock. Semi-regular oscillations begin only after
the density contrast has become high enough for an efficient shock
reflection.
\end{enumerate}
\subsection{Dependence of the evolution on shock velocity}
						\label{s:evol_v}
In both cases described above the evolution starts with a nearly
adiabatic expansion of the main shock followed by formation of the thin
shell. In the case of the fast shock ($v_{\rm s0}=6000$ km s$^{-1}$,
Figs.\ \ref{f:den}(b), \ref{f:time_dyn}(b), and \ref{f:time_lum}b)
%
%
\begin{figure*}
  \centering
  \vspace*{15pc}
  \caption{
    Temporal variations of total, HE, ME, and LE luminosity of the post-shock
region
    (thick, medium, thin and dotted lines, respectively).
(a) initial shock velocity $v_{\rm s0}=2000$ km s$^{-1}$;
    luminosity scale: $3.71\times10^{7}$ erg cm$^{-2}$s$^{-1}$.
(b) initial shock velocity $v_{\rm s0}=6000$ km s$^{-1}$;
    luminosity scale: $1.52\times10^{8}$ erg cm$^{-2}$s$^{-1}$.
}
  \label{f:time_lum}
\end{figure*}
the shell condenses far away from the rigid left boundary of the grid,
and it moves outwards due to high pressure in the hot, rarefied gas left
behind it. This region (called {\em hot cavity} by Terlevich et al.\
1992) is created as the post-shock temperature drops faster due to
rapidly decreasing inflow velocity than due to cooling (as a result,
catastrophic cooling occurs closer to the shock). This is not true for
shocks slower than $\sim 3000$ km s$^{-1}$ as the initial post-shock
temperatures are lower by an order of magnitude and, therefore, the
cooling is much more effective. In that case the shell recedes to the
left, and we observe the formation of the second shell at the left
boundary of the grid. The collision of two shells at the end of the run
for slower shock (upper left part of Fig.\ \ref{f:den}a) is marked by
prominent luminosity spike at time $t\approx 2.5$ yr (Fig.\
\ref{f:time_lum}a). Finally even for shocks of higher velocities the
shell will collide with the left boundary of the grid.
\section{Photometric and spectral features}
						\label{s:ph_and_spec}
\subsection{Broad band emission}
						\label{s:broad_em}
The behavior of the broad band emission of ionizing radiation produced
in the post-shock region is shown in Fig.\ \ref{f:time_lum}. During the
first evolutionary phase associated with nearly adiabatic shock
expansion the total luminosity smoothly increases, and the shock slows
down. At the moment of shell formation LE/ME luminosities rapidly
increase while HE emission remains nearly constant. Fast variability of
LE/ME emission during shell formation indicates nonuniformity of the gas
entering the shell. This nonuniformity of the cool, condensing gas arises
from small, numerical fluctuations present in the post-shock region of
nearly standing main shock (see Fig.\ \ref{f:den}) which are amplified
by catastrophic cooling.

After shell formation (at $t\approx 1$ and $t\approx3.2 $ yr for slow
and fast shock, respectively) the HE luminosity steadily declines. The
decline rate is higher for the slow shock, as in this case the hot
region behind the shell has lower temperature. In the case of fast shock
this hot, rarefied gas radiates the rest of its thermal energy at
$t\approx 10.5$ yr.

The main shock reflection from the shell ($t\approx 1.6$ yr, Figs.
\ref{f:time_dyn}(a) and \ref{f:time_lum}(a); $t\approx 5.0$ yr, Figs.
\ref{f:time_dyn}(b) and \ref{f:time_lum}b) is marked by the kink in
shock position and an intense burst of the LE/ME radiation. From now on,
LE and ME luminosities begin to oscillate on a time scale roughly
corresponding to the fundamental mode of the oscillatory instability
(Sect.\ \ref{s:evol_phases}). Note that the post-shock region of the
oscillating shock does not radiate in HE, therefore, no rapid variations
of the high energy emission are expected.

For reasons explained in Sect.\ \ref{s:evol_phases} in the next two
sections we will concentrate our discussion on the spectrum emitted by
the shock of original velocity $v_{\rm s0}=6000$ km s$^{-1}$ (roughly
equal to the velocity of the forward shock considered by Terlevich et
al.\ 1992).
\subsection{The continuum}
						\label{s:contin}
The emergent shock spectrum just after the thin shell formation (at
$t\approx 3.20$ yr, i.e.\ at the moment the shock achieves its maximum
luminosity) is shown in Fig.\ \ref{f:spec6}.
%
%
\begin{figure*}
  \centering
  \vspace*{15pc}
  \caption{
    The emergent spectrum of the shock 6000 km s$^{-1}$ at time $t=3.20$
    yr, just after the thin shell formation.
    (a) emergent flux;
    (b) transmitted flux;
    (c) solid line: sum of the reflected flux and half
    of the flux generated between the main shock and the shell;
        dashed: reflected flux only.
    The energy flux $\nu F_{\nu}$ (erg cm$^{-2}$s$^{-1}$) is plotted in
    logarithmic scale. Bottom and top scales are given in logarithm of
    energy in Rydbergs and keV, respectively.
    See text for details.}
  \label{f:spec6}
\end{figure*}
The energy flux $\nu F_{\nu}$ (Fig.\ \ref{f:spec6}a) is the sum of three
components (Figs.\ \ref{f:spec6}(b) and \ref{f:spec6}c):
\begin{enumerate}[{3}]
\renewcommand{\theenumi}{(\arabic{enumi})}
\item the transmitted flux (originating between the shell and the
left edge of the grid, and partly absorbed by the shell);
\item the half of the flux generated between the main shock and the shell
(emitted directly towards the observer);
\item the reflected part of the second half of flux generated between the
main shock and the shell (originally emitted towards the shell and
reflected by the shell towards the observer).
\end{enumerate}
The emergent spectrum has a power-law shape
$F_{\nu}\propto\nu^{-\alpha}$ with a mean spectral index $\alpha\approx
0.63$, and with a high-energy cutoff around 10 keV. For energies above
few keV the spectrum is completely dominated by the emission produced by
the hot region between the shell and the left boundary of the grid that
has not yet had enough time to cool down due to its very high initial
temperature (see Sections \ref{s:evol_v} and \ref{s:broad_em}). This
part of the spectrum is unlikely to be affected by any improvement of
cooling function for nonequilibrium effects. For lower energies, most of
the radiation comes from the post-main-shock region located between the
main shock and the shell. The UV/optical emission is produced by the
shell ionized predominantly by photons emitted from the hot region
obscured by the shell. As the emission from this region is modeled
correctly, the low-energy part of the shock spectrum is also calculated
correctly.
\subsection{The line emission}
The ionizing flux, as defined in Sect.\ \ref{s:method}, consists of two
components only:
\begin{enumerate}[{3}]
\renewcommand{\theenumi}{(\arabic{enumi})}
\item the flux originating between the shell and the left edge of the
grid;
\item half of the flux generated between the main shock and the shell.
\end{enumerate}
The ionized medium was identified with the dense shell. The shell of the
6000 km s$^{-1}$ shock contained $\sim50$ zones and its average
thickness was $\sim4\times10^{13}$ cm. At selected evolutionary
times the ionizing spectra together with corresponding shell models were
used as the input to CLOUDY photoionization code.

The output from CLOUDY consisted of intensities and intensity ratios of
selected lines as functions of time. It was found that due to finite
time of the shell formation line intensity maxima lagged behind ionizing
flux maximum, the lag being shorter for high ionization lines, and
longer for low ionization ones (Table \ref{t:lag}).
%
%
\begin{table}
  \centering
  \caption{Time lags for selected emission lines.}
  \label{t:lag}
  \begin{tabular}{lccc}
  \hline \hline
       line                      & \multicolumn{3}{c}{time lag [d]} \\
                                 & observed & TTRFM  & this paper   \\
  \hline
       H$\beta$ $\lambda 4861$   &     20   & 30-40  & 17.9         \\
       Ly$\alpha$ $\lambda 1216$ &     12   &  1-15  &  5.1         \\
       He II $\lambda 1640$     &   4-10   & 12-20  &  5.1         \\
       Mg II $\lambda 2798$     &  34-72   & 40-50  & 40.5         \\
       C III] $\lambda 1909$    &  26-32   & 20-30  & 25.2         \\
       Si IV $\lambda 1397$     &  12-34   & 12-17  &  5.8         \\
       C IV $\lambda 1549$      &   8-16   & 10-15  &  3.3         \\
       N V $\lambda 1240$       &      4   &  2-10  &  5.1         \\
  \hline
  \end{tabular}
\end{table}
Temporal variations of selected line intensities and ionizing flux
intensities during the shell formation phase are shown in Fig.\
\ref{f:lag6}.
%
%
\begin{figure*}
  \centering
  \vspace*{25pc}
  \caption
{
Temporal variations of selected line intensities and ionizing flux during
formation of the shell for initial shock velocity $v_{\rm s0}=6000$ km
s$^{-1}$. All values are shown in logarithmic scale; time is given in
years on horizontal axis. Vertical line marks the time of maximum
intensity of ionizing flux.
}
  \label{f:lag6}
\end{figure*}
Fig.\ \ref{f:photo7}
%
%
\begin{figure*}
  \centering
  \vspace*{30pc}
  \caption
{
Temporal variations of selected line ratios and parameters derived from
photoionization models for initial shock velocity $v_{\rm
s0}=6000$ km s$^{-1}$. Left scale: full line; right scale: dotted line.
Line ratios typical for AGNs are marked by dots on vertical axes. All
values are shown in logarithmic scale; time is given in years on
horizontal axis.
}
  \label{f:photo7}
\end{figure*}
presents temporal variations of selected line ratios throughout the
simulated evolution together with variations of the total hydrogen column
density $N({\rm H})$ and maximum density of the shell $n_{\rm max}$.
\section{Discussion}
						\label{s:discussion}
The general conclusion of our calculations is that  the basic features
of the fast shock model are similar to those observed in AGNs.

First, the emergent spectrum of the shock has a power-law shape
$F_{\nu}\propto\nu^{-\alpha}$ with a mean spectral index of
$\alpha\approx 0.63$, and with a high-energy cutoff around 10 keV. Note
that the hard X-ray ($2-10$ keV) AGN spectrum is characterized by a
single power-law with $\alpha=0.7$ (Turner \& Pounds 1989). The high
energy end of the spectrum (hereafter the cutoff energy) is produced by
the gas located behind the shell, and we recall that this gas was
processed at the beginning of evolution by the shock moving with a
velocity nearly equal to the original shock velocity. Assuming that the
cooling was ineffective and the temperature of the gas did not change
appreciately till the time of the shell formation we can state that the
cutoff energy is directly related to the original shock velocity.
Therefore, one may expect that the spectrum of the shock with the
initial velocity equal to the initial velocity of a supernova blast wave
($v_{\rm s0}=1-2\times10^{4}$ km s$^{-1}$) would have the energy cutoff
at around 100 keV, as the post-shock temperature is proportional to
square of the shock velocity (such models were not obtained as we
considered evolution for $t\ge t_{\rm sg}$, see Sect.\
\ref{s:dyn_evol}). This is the cutoff value expected for Seyfert
galaxies (see Zdziarski, \.Zycki \& Krolik (1993b) for references). Note
also that our model shock spectrum very closely resembles thermal
Comptonization component used in calculations of X-ray spectra of AGNs
(Zdziarski et al. 1993b, Zdziarski, Lightman \& Macio\l ek-Nied\'zwiecki
1993a).

Second, intensity ratios of spectral lines produced in the photoionized
shell agree fairly well with those measured in AGNs (Fig.\
\ref{f:photo7}). The values for AGNs in Fig.\ \ref{f:photo7} were taken
after compilation of Kwan \& Krolik \shortcite{KK81}, and they are
marked on the vertical axes by dots. We also confirm preliminary results
obtained by Terlevich et al.\ \shortcite{ACEROSI}. The predicted total
hydrogen column densities of the ionized medium are of the order of
$10^{23}$ cm$^{-2}$ and the maximum gas density is roughly equal to
$10^{11}$ cm$^{-3}$. Our model fails in explanation of low
Ly$\alpha$/H$\beta$ ratio, but this problem is common for other
photoionization models of AGNs.

Finally, our model exhibits {\em time lag}, i.e.\ a delay between maxima
of ionizing continuum intensity and line intensities. TTRFM using a very
simple model of the shell and ionizing continuum obtained time lags
listed in Table \ref{t:lag}. With our study, which is entirely free of
arbitrary assumptions concerning the shell and the ionizing continuum,
we corroborate their findings. The predicted time lags agree well with
those observed in AGNs. However, this result should be treated with
some caution as the observational definition of the time lag (involving
statistical measurements) is different from the one employed here.

Taken together, the above properties provide an indication that cSNRs
play a significant role in the AGN phenomenon, as it was proposed in the
starburst model (see, e.g., Terlevich et al.\ 1992). One may be
almost sure that more sophisticated model spectra based on appropriate
assemblage of shocks with different velocities would recover the ME
excess observed in AGNs (i.e.\ the so called big blue bump, Turner \&
Pounds 1989). This is because in a massive starburst there will
be many cSNRs coexisting, most of them in relatively long lasting
advanced evolutionary phases characterized by low shock velocities. Such
models would also enable statistical measurements of time lags better
suited for comparison with observations than these presented here.

Our conclusion is that cSNR shocks are an attractive research field
whose exploration has only begun here. The finding that time lag
phenomena may be related to variations of physical parameters of the
photoionized medium (as originally proposed by TTRFM) seems to be
particularly attractive and it clearly deserves future investigation.
\subsection*{Acknowledgments}
I thank Micha{\l } R\'o\.zyczka for his steady support and many valuable
discussions, and Gary Ferland for making his code available to me.
Thanks are also due to anonymous referee whose comments helped to
improve and clarify the paper. This work was supported by the Committee
of Scientific Research through the grant 2-1213-91-01 and by the ESO
C\&EE grant A-01-063.
\bsp
\end{document}